\documentstyle[12pt]{article}

\catcode`\@=11
\def\section{\@startsection{section}{1}{0.0ex}
                   {3.5ex plus -1.0ex minus -0.2ex}
                   {2.3ex plus 0.2ex}{\bf}}
\def\subsection{\@startsection{subsection}{2}{0.0ex}
                        {3.25ex plus 1ex minus .2ex}
                        {1.5ex plus .2ex}{\bf}}

\catcode`@=12
\newskip\humongous \humongous=0pt plus 1000pt minus 1000pt
\def\caja{\mathsurround=0pt}

\newif\ifdtup
\def\panorama{\global\dtuptrue \openup1\jot \caja
        \everycr{\noalign{\ifdtup \global\dtupfalse
        \vskip-\lineskiplimit \vskip\normallineskiplimit
        \else \penalty\interdisplaylinepenalty \fi}}}
\def\eqalignno#1{\panorama \tabskip=\humongous
        \halign to\displaywidth{\hfil$\displaystyle{##}$
        \tabskip=0pt&$\displaystyle{{}##}$\hfil
        \tabskip=\humongous&\llap{$##$}\tabskip=0pt
        \crcr#1\crcr}}
\begin{document}
\def\thefootnote{\fnsymbol{footnote}}
\begin{titlepage}
\vspace*{-0.7in}
\begin{center}
{\bf A Non-Minimal {\mbox{$\bf SO(10) \times U(1)_F$}} SUSY - GUT Model\\
        Obtained from a Bottom-Up Approach} \\
\vskip 0.20in
        {\bf Carl H. ALBRIGHT}\\
 Department of Physics, Northern Illinois University, DeKalb, Illinois 
60115\footnote{Permanent address}\\[-0.2cm]
        and\\[-0.2cm]
 Fermi National Accelerator Laboratory, P.O. Box 500, Batavia, Illinois 
60510-0500\footnote{Electronic address: albright@fnal.gov}\\
\end{center}
\vfill
\begin{abstract}
\indent  Many of the ingredients are explored which are needed to develop 
a 
supersymmetric $SO(10) \times U(1)_F$ grand unified model based on the 
Yukawa 
structure of a model previously constructed in collaboration with S. 
Nandi to 
explain the quark and lepton masses and mixings in a particular neutrino 
scenario.  The $U(1)_F$ family symmetry can be made anomaly-free with the 
introduction of one conjugate pair of $SO(10)$-singlet neutrinos with the 
same
$U(1)_F$ charge.  Due to a plethora of conjugate pairs of 
supermultiplets, the 
model develops a Landau singularity within a factor of 1.5 above the GUT 
scale.
With the imposition of a $Z_2$ discrete symmetry and under certain 
conditions,
all higgsino triplets can be made superheavy while just one pair of 
higgsino 
doublets remains light and results in mass matrix textures previously 
obtained 
from the bottom-up approach.  Diametrically opposite splitting of the 
first 
and third family scalar quark and lepton masses away from the second 
family 
ones results from the nonuniversal D-term contributions.
\end{abstract}
\noindent PACS numbers: 12.15.Ff, 12.60Jv\hfill <e-Print Archive: 
        hep-ph/9608372>
\end{titlepage}
\setcounter{page}{2}
\section{INTRODUCTION}
In a recent series of papers \cite {an1,an2}, the author in 
collaboration with S. Nandi
began a program to construct a viable model for the fermion quark and 
lepton masses and mixings at the supersymmetric grand unification scale.  
The program envisaged by us has evolved in three stages, beginning with a 
bottom-up approach which ensures accurate results for the known 
low-energy 
data without introducing an undue amount of theoretical bias at the 
outset. 
This is to be contrasted with most theoretical model construction which 
has 
been carried out by various authors \cite{top-down} using a top-down 
approach.  In that case, some well-defined theoretical principles are 
selected at the outset with the 
model parameters then picked to fit the known low-energy data as well as 
possible.

The general framework chosen by us was that of supersymmetric $SO(10)$ 
grand unification (SUSY-GUTS), since this appeared to give the most 
satisfactory explanation for the unification of the standard model gauge 
couplings \cite{guni} at a high energy scale of the order of $10^{16}$ 
GeV, 
as well as accommodating the 16 fermions of each family in a simple 
representation of the gauge group.  The low energy data for
most quark and charged lepton masses as well as the quark 
Cabibbo-Kobayashi-Maskawa quark mixing matrix \cite{CKM} are reasonably 
well-known \cite{PDB}, while various scenarios must be entertained at this
time for the neutrino masses and mixings according to which experimental
results one is willing to accept at face value.

The first bottom-up stage \cite{an1} of our program for a given scenario 
then consisted of 
evolving \cite{Nac} the masses and mixing matrices to the SUSY-GUT scale, 
where the up, down, charged lepton and neutrino mass matrices can be 
constructed by making use of Sylvester's theorem as illustrated originally
for quark mass matrices by Kusenko \cite{Syl}.  Two free 
parameters, one for the 
quark sector and one for the lepton sector, which control the choice of 
bases for the mass matrices were tuned and different neutrino scenarios
selected to search for mass matrices exhibiting simple $SO(10)$ structure.
For this purpose, complete unification of all third family
quark and lepton Yukawa couplings was assumed \cite{uni} corresponding to 
a pure 
$\bf {10}$ Higgs contribution to the 33 mass matrix elements, while 
simplicity in the sense of pure $\bf {10}$ or pure $\bf {\overline{126}}$ 
Higgs contributions was sought for as many of the other mass matrix 
elements as possible. This choice of procedure was influenced by earlier 
work such as that of Georgi and Jarlskog \cite{GJ}, where the 33 mass 
matrix elements transformed as pure ${\bf 10}$'s and the 22 elements as 
pure ${\bf 126}$'s.  
We are aware that level-5 $\bf \overline{126}$ 
$SO(10)$ multiplets do not arise naturally in superstring models \cite{126}
and must be treated as effective operators; hence such a model should be 
treated as an effective theory at best. We shall return to this point 
at the end of Sect. II.

The simplest $SO(10)$ structure at the SUSY-GUT scale was obtained in the 
neutrino scenario incorporating the Mikheyev - Smirnov - Wolfenstein (MSW)
\cite{MSW} nonadiabatic resonant 
conversion interpretation of the depletion of solar electron-neutrinos 
\cite{solar} together with the observed depletion of atmospheric 
muon-neutrinos \cite{atm}.  In this scenario,
no eV-scale neutrino masses exist to contribute a hot dark matter component
to mixed dark matter \cite{MDM}; moreover, since no sterile neutrinos 
were 
incorporated into the model at that time, in the version under consideration
we are unable to explain the $\bar{\nu}_{\mu} \rightarrow
\bar{\nu}_e$ mixing results obtained by the LSND collaboration \cite{LSND}.
The mass 
matrices constructed at the SUSY-GUT scale have the following textures
        $$M^U \sim M^{N_{Dirac}} \sim diag(\overline{126};\ \overline{126};
                \ 10)\eqno(1.1a)$$
        $$M^D \sim M^E \sim \left(\matrix{10',\overline{126} & 10',
                \overline{126}' & 10'\cr 10',\overline{126}' & 
\overline{126} 
                & 10'\cr 10' & 10' & 10\cr}\right)\eqno(1.1b)$$
with $M^D_{11},\ M^E_{12}$ and $M^E_{21}$ anomalously small and only the 
13 and 31 elements complex.    Entries in the matrices stand for the Higgs
representations contributing to those elements.
We assumed that vacuum expectation values (VEV's) 
develop only for the symmetric representations ${\bf 10}$ and ${\bf 126}$.
The ${\bf 10}$'s contribute equally to
$(M^U,\ M^D)$ and $(M^{N_{Dirac}},\ M^E)$, while the ${\bf 126}$'s
weight $(M^U,\ M^D)$ and $(M^{N_{Dirac}},\ M^E)$ in the ratio of 
1\ : -3.  The Majorana neutrino mass matrix $M^R$, 
determined from the seesaw formula \cite{GMRS} with use of $M^{N_{Dirac}}$
and the 
reconstructed light neutrino mass matrix, exhibits a nearly geometrical 
structure \cite{geom} given by 
$$M^R \sim \left(\matrix{F & - \sqrt{FE} & \sqrt{FC}\cr
                - \sqrt{FE} & E & -\sqrt{EC}\cr
                \sqrt{FC} & -\sqrt{EC} & C\cr}\right)
                        \eqno(1.1c)$$
where $E \simeq {5\over{6}}\sqrt{FC}$ with all elements relatively real.  
It can not be purely geometrical, however, since the singular rank-1 
matrix 
above can not be inverted as required by the seesaw formula, $M^{N_{eff}} 
\simeq -M^{N_{Dirac}}(M^R)^{-1} M^{N^T_{Dirac}}$.

In the second stage \cite{an2} of our program, we attempted to find a 
model incorporating a family symmetry which yields the above matrices 
determined phenomenologically
from our bottom-up approach.  Success was obtained by introducing a global
$U(1)_F$ family symmetry \cite{U1} which uniquely labels each one of the 
three light
families, as well as conjugate pairs of heavy families and various Higgs 
representations.  In addition to controlling the evolution of the Yukawa
couplings from the SUSY-GUT scale to the supersymmetry-breaking weak scale,
the supersymmetric nature of the SUSY-GUT model played a 
key role in that the nonrenormalization theorems \cite{nonren} of 
supersymmetry allow
one to focus solely on Dimopoulos-type tree diagrams \cite{Dim}, in order 
to calculate 
the contributions to the mass matrix elements.  With twelve input parameters
in the form of Yukawa couplings times VEV's, the numerical results 
obtained for the 3 heavy Majorana masses and 20 low energy parameters for 
the quark and lepton masses and two mixing matrices 
were found to be in exceptionally good agreement with the low energy data
in the neutrino scenario in question as shown in \cite{an2}.

In this paper the author has pursued the third stage of the program which 
is 
an attempt to construct a consistent supersymmetric grand unified model 
of all 
the interactions in an $SO(10) \times U(1)_F$ framework.  A number of 
important issues \cite{SUSY} must be addressed such as the anomaly-free 
nature 
of the 
superpotential, the requirement that supersymmetry remain unbroken at the 
SUSY-GUT scale and only effectively broken at the electroweak
scale, the requirement that all colored Higgs triplets become superheavy,
so that proton decay remains sufficiently suppressed while only one pair 
of 
Higgs doublets remains light to break the electroweak 
symmetry and to preserve
the good prediction for the electroweak angle in $\sin^2 \theta_W$.  The
evolution equations for the gauge and Yukawa couplings should also not be 
greatly altered by the presence of any extra light fields in the model.  
In 
this paper we shall explore some of these issues and present many of the 
ingredients needed to construct such a model; however, due to the 
complexity 
of the model we have not checked that all conditions can be imposed in a 
self-consistent fashion to achieve the desired VEV's and to avoid 
problems 
with the low energy phenomenology.  We find that a Landau singularity 
develops 
slightly above the SUSY-GUT scale as a result of the plethora of 
conjugate 
pairs of supermultiplets present in the model.  Some indication of the
type of splitting of the squark and slepton masses is also gained from
the nonuniversal D-term contributions.

\section{SUPERMULTIPLETS AND HIGGS SUPERPOTENTIAL}

In order to build a satisfactory supersymmetric model \cite{SUSY}, we 
require 
that 
the superpotential be analytic and anomaly-free.  For this purpose we
replace the fermion and Higgs $SO(10)$ multiplets by chiral supermultiplets
and double the number of Higgs supermultiplets with non-zero $U(1)_F$ charges
by adding supermultiplets with equal and opposite $U(1)_F$ charges.

We begin by listing the fermion and boson fields introduced in the 
$SO(10) \times U(1)_F$ model in the second stage of our program.\\
\newpage
\indent {\bf $\bullet$ Left-Handed Fermion Fields:}
        $${\bf 16}_i: \qquad \psi^{(9)}_3,\quad \psi^{(-1)}_2, \quad 
                \psi^{(-8)}_1 \eqno(2.1a)$$
        $$\begin{array}{lrrrrrrrrrrrrr}
  {\bf 16}: & f:&-0.5,&1.0,&2.0,&4.0,&4.5,&-4.5,&-7.5,&11.0,&12.5,\nonumber\\
        & & & & & 1.5,&-6.0,&-6.5 \\
  {\bf \overline{16}}: & f^c:&0.5,&-1.0,&-2.0,&-4.0,&-4.5,&4.5,&7.5,&
        -11.0,&-12.5, \\
        & & & & & -1.5,&6.0,&6.5 \cr \end{array} \eqno(2.1b)$$
We have identified with a subscript the three light fermion family fields 
belonging to the ${\bf 16}$ representations of $SO(10)$ and indicated
their assigned $U(1)_F$ charges with a superscript, while for the 
conjugate superheavy fermion fields we have just listed their $U(1)_F$ 
charges. The corresponding Higgs boson fields comprise the following:\\
\indent {\bf $\bullet$ Higgs Fields}:
        $$\begin{array}{rl}
        {\bf 10}:& H^{(-18)}_1,\ H^{(-8)}_2 \nonumber\\
        {\bf 45}:& A^{(3.5)}_1,\ A^{(0.5)}_2 \\
        {\bf \overline{126}}:& \bar{\Delta}^{(2)},\ \bar{\Delta}'^{(-22)} \\
        {\bf 1}:& S^{(2)}_1,\ S^{(6.5)}_2 \cr \end{array} \eqno(2.1c)$$

As customary, for each of the above fields we introduce a chiral 
superfield 
with the same $U(1)_F$ charge and components as indicated:
        $$\begin{array}{lclrl}
        {\bf \Psi}_i &=& (\tilde{\psi}_i,\ \psi_i,\ \chi_{\psi_i}), 
                \qquad &i &= 1,2,3 \nonumber\\
        {\bf F}_i &=& (\tilde{f}_i,\ f_i,\ \chi_{f_i}), \qquad &i &= 1 - 
12 \\
        {\bf \bar{F}}_i &=& (\tilde{f}^c_i,\ f^c_i,\ \chi_{f^c_i}) 
                \qquad &i &= 1 - 12 \\
        {\bf H}_i &=& (H_i,\ \tilde{H}_i,\ \chi_{H_i}), \qquad &i &= 1,2 \\
        {\bf A}_i &=& (A_i,\ \tilde{A}_i,\ \chi_{A_i}), \qquad &i &= 1,2 \\
        {\bf \bar{\Delta}} &=& (\bar{\Delta},\ \tilde{\bar{
                \Delta}},\ \chi_{\bar{\Delta}}), \qquad 
        & {\bf \bar{\Delta}'} &= (\bar{\Delta}',\ \tilde{\bar{
                \Delta}}',\ \chi_{\bar{\Delta}'}) \\
        {\bf S}_i &=& (S_i,\ \tilde{S}_i,\ \chi_{S_i}), \qquad &i &= 1,2 
                \cr \end{array} \eqno(2.2)$$
All chiral superfields are taken to be left-handed $SO(10)$ supermultiplets;
the tildes indicate superpartners of the ordinary fermions or bosons with 
odd R-parity; and the $\chi's$ refer to the corresponding auxilary fields.

In order that the superpotential to be constructed will be analytic and 
anomaly-free,
we double the superfields containing the ordinary Higgs scalars by 
introducing
superfields with the opposite $U(1)_F$ charges and conjugate $SO(10)$
representations:
        $$\begin{array}{lclrl}
        {\bf \bar{H}}_i &=& (\bar{H}_i,\ \tilde{\bar{H}}_i,\ 
                \chi_{\bar{H}_i}), \qquad &i &= 1,2 \\
        {\bf \bar{A}}_i &=& (\bar{A}_i,\ \tilde{\bar{A}}_i,\ 
                \chi_{\bar{A}_i}), \qquad &i &= 1,2 \\
        {\bf \Delta} &=& (\Delta,\ \tilde{\Delta},\ \chi_{\Delta}), \qquad
        & {\bf \Delta'} &= (\Delta',\ \tilde{\Delta}',\ \chi_{\Delta'}) \\
        {\bf \bar{S}}_i &=& (\bar{S}_i,\ \tilde{\bar{S}}_i,\ 
                \chi_{\bar{S}_i}), \qquad &i &= 1,2 \cr \end{array}
                \eqno(2.3)$$
Since the sum of the $U(1)_F$ charges for the three light fermion families
is zero, the $[SO(10)]^2 \times U(1)_F$ triangle anomaly vanishes.  
The remaining $[U(1)_F]^3$ triangle anomaly can be canceled with the 
introduction of just two singlet (sterile) neutrinos, $n$ and $n^c$, both 
with $U(1)_F$ charge of -12 which prevents them from pairing off and 
becoming superheavy \cite{anomcanc}.  The corresponding superfields are 
        $$\begin{array}{lcl}
        {\bf N} &=& (\tilde{n},\ n,\ \chi_n) \nonumber\\
        {\bf {\bar{N}}} &=& (\tilde{n}^c,\ n^c,\ \chi_{n^c}) \cr
                \end{array} \eqno(2.4)$$

In addition to the analyticity and anomaly-free requirements for the 
superpotential, we must ensure that many fields become superheavy at 
the SUSY-GUT scale $\Lambda_{SGUT}$, while three fermion families of 
${\bf 16}$'s remain light.  Moreover, just one pair of Higgs doublets 
should remain light \cite{1pair} to ensure a good value for $\sin^2 
\theta_W$,
while all colored Higgs triplets must get superheavy to avoid rapid 
proton 
decay via dimension 5 and 6 operators \cite{colHiggs}.  This can be 
accomplished by introducing
some additional chiral Higgs superfields transforming as $SO(10)$ 
representations which do not participate in the Yukawa interactions for 
which the original $SO(10) \times U(1)_F$ model was constructed.

To help identify a suitable choice of additional superfields, we elaborate
the maximal $SU(2)_L \times SU(2)_R \times SU(4)$ subgroup content of 
various 
$SO(10)$ representations \cite{Slan}.
        $$\begin{array}{lrcl}
          {\bf H}:&\qquad {\bf 10} &=& (2,2,1) + (1,1,6) \nonumber\\
          {\bf A}:&     {\bf 45} &=& (1,1,15) + (1,3,1) + (3,1,1) + 
(2,2,6) \\
          {\bf \Sigma}:&  {\bf 54} &=& (1,1,1) + (3,3,1) + (1,1,20') + 
                (2,2,6) \\
          {\bf \Delta}:&  {\bf 126} &=& (1,3,\bar{10}) + (3,1,10) +
                (1,1,6) + (2,2,15) \\
          {\bf \bar{\Delta}}:&  {\bf \overline{126}} &=& (1,3,10) + 
                (3,1,\overline{10}) + (1,1,6) + (2,2,15) \\
          {\bf \Phi}:&  {\bf 210} &=& (1,1,1) + (1,1,15) + (1,3,15) +
                (3,1,15) \\
                & & & \qquad + (2,2,6) + (2,2,10) + (2,2,\overline{10}) \\
        \end{array} \eqno(2.5)$$
This suggests that $\Lambda_{SGUT}$ scale VEV's can be generated for each
$SO(10)$ representation while preserving the standard model gauge group
according to
        $$\begin{array}{rlcl}
          {\bf 1}:& \qquad <S> &=& ts_{1,1,1} \nonumber\\
          {\bf 45}:& \qquad <A> &=& pa_{1,1,15} + qa_{1,3,1} \\
          {\bf 54}:& \qquad <\Sigma> &=& r\sigma_{1,1,1} \\
          {\bf 126}:& \qquad <\Delta> &=& v_R\delta_{1,3,\overline{10}} \\
          {\bf \overline{126}}:& \qquad <\bar{\Delta}> &=& 
                {\bar{v}}_R\delta_{1,3,10} \\
          {\bf 210}:& \qquad <\Phi> &=& a\phi_{1,1,1} + b\phi_{1,1,15}
                + c\phi_{1,3,15} \\
                \end{array} \eqno(2.6)$$
where the VEV directions in the $SU(2)_L \times SU(2)_R \times SU(4)$
subspace follow from (2.5) and the coefficients are in general complex.

Higgsino doublets containing a neutral field are generated when the $SO(10)$
representations break down to the standard model (SM) 
$SU(3)_c \times SU(2)_L \times U(1)_Y$ gauge group according to 
        $$\begin{array}{rclclcllcl}
          {\bf 10} &\supset& (2,2,1) &\supset& \tilde{H}_u &=& \left(
                \matrix{h^+ \cr h^o \cr}\right),\quad & \tilde{H}_d &=& 
                \left(\matrix{\bar{h}^o \cr h^- \cr}\right)
                \nonumber\\[0.2in]
          {\bf \overline{126}} &\supset& (2,2,15) &\supset& \tilde{
                \bar{\Delta}}_u &=& \left(
                \matrix{\delta^+ \cr \delta^o \cr}\right),\quad & \tilde{
                \bar{\Delta}}_d &=& 
                \left(\matrix{\bar{\delta}^o \cr \delta^- \cr}\right)
                \nonumber\\[0.2in] 
          {\bf 210} &\supset& (2,2,\overline{10}) &\supset& \tilde{
                \Phi}_u &=& \left(\matrix{\phi^+ \cr \phi^o \cr}\right) 
                \\[0.2in]
          {\bf 210} &\supset& (2,2,10) &\supset& \tilde{\Phi}_d &=&
                \left(\matrix{\bar{\phi}^o \cr \phi^- \cr}\right) \\
        \end{array} \eqno(2.7)$$
Electroweak scale VEV's are generated by Higgs scalars in the ${\bf 10}$ and
${\bf \overline{126}}$ superfields when the standard model breaks to 
$U(1)_{em}$ and can give masses to the three families of fermions in the 
$\psi_i$ of (2.1a) according to
        $$\begin{array}{rlcl}
          {\bf 10}:& \qquad <H> &=& v_uh_{2,1,1} + v_dh_{2,-1,1} \nonumber\\
          {\bf \overline{126}}:& \qquad <\bar{\Delta}> &=&  
                w_u\delta_{2,1,1} + w_d\delta_{2,-1,1} \\
          \end{array} \eqno(2.8)$$
Here the subscripts refer to the VEV directions in the $SU(2)_L \times U(1)_Y
\times SU(3)_c$ basis.
As noted earlier, just one pair of Higgs doublets should remain light at 
the electroweak scale, so a good value for $\sin^2 \theta_W$ is obtained.
How this can come about is discussed in detail in Sect. III.

Higgsino colored triplets of charges $\pm 1/3$ which can couple to a pair 
of 
quarks and a quark and lepton and hence be exchanged in a diagram leading 
to 
proton decay appear in
        $$\begin{array}{rclclcllcl}
        {\bf 10} &\supset& (1,1,6) &\supset& \tilde{H}_t &=& 
                h^{-1/3}, \quad & \tilde{H}_{\bar{t}} &=& h^{1/3} 
                \nonumber\\
        {\bf 126} &\supset& (1,1,6) &\supset& \tilde{\Delta}^{(1,1,6)}_t 
&=& 
                \delta^{-1/3}, \quad & \tilde{\Delta}^{(1,1,6)}_{\bar{t}} 
&=& 
                \delta^{1/3} \\
        {\bf 126} &\supset& (1,3,\overline{10}) &\supset& \tilde{\Delta}^
                {(1,3,\overline{10})}_{\bar{t}} &=& \delta'^{1/3} \\
        {\bf \overline{126}} &\supset& (1,1,6) &\supset& 
                \tilde{\bar{\Delta}}^{(1,1,6)}_t &=& \bar{\delta}
                ^{-1/3}, \quad & \tilde{\bar{\Delta}}^{(1,1,6)}_{\bar{t}}
                &=& \bar{\delta}^{1/3} \nonumber\\
        {\bf \overline{126}} &\supset& (1,3,10) &\supset&
                \tilde{\bar{\Delta}}^{(1,3,10)}_t &=& \bar{\delta}'^
                {-1/3} \\
        {\bf 210} &\supset& (1,3,15) &\supset& \tilde{\Phi}_t &=&
                \phi^{-1/3}, \quad & \tilde{\Phi}_{\bar{t}} &=&
                \phi^{1/3} \\
        \end{array} \eqno(2.9)$$
We shall discuss the issue of surviving light Higgs triplets in Sect. IV.

In order to generate a satisfactory higgsino doublet mass matrix, we find
it necessary to add the following Higgs superfields:
        $${\bf 54}:\quad {\bf \Sigma}^{(0)}_0,\ {\bf \Sigma}^{(-16)}_1,\ 
{\bf 
                \bar{\Sigma}}^{(16)}_1,\ {\bf \Sigma}^{(-10)}_2,
                \ {\bf \bar{\Sigma}}^{(10)}_2 \eqno(2.10a)$$
\vspace{-0.2in}
        $${\bf 210}:\quad {\bf \Phi}^{(0)}_0,\ {\bf \Phi}^{(-20)}_1,\ 
{\bf 
                \bar{\Phi}}^{(20)}_1,\ {\bf \Phi}^{(-10)}_2,
                \ {\bf \bar{\Phi}}^{(10)}_2 \eqno(2.10b)$$
To make all the higgsino triplets of type (2.9) superheavy, we introduce
the additional Higgs superfield:
        $${\bf 45}: \quad {\bf A}^{(0)}_0  \eqno(2.10c)$$
Finally, we must introduce the following Higgs superfields to guarantee
F-flat directions, so supersymmetry is only softly broken at $\Lambda_{SGUT}$
as discussed in Sect. V.
        $$\begin{array}{rl}
          {\bf 45}:&\quad {\bf A}^{(8)}_3,\ {\bf \bar{A}}^{(-8)}_3 
\nonumber\\
          {\bf 1}:&\quad {\bf S}^{(8.5)}_3,\ {\bf \bar{S}}^{(-8.5)}_3 \\
          \end{array} \eqno(2.10d)$$

We are now in a position to write down all the terms in the 
superpotential 
which conserve the $U(1)_F$ charge.  The Higgs superpotential for the 
quadratic and cubic terms is given by
        $$\begin{array}{rcl}
          W^{(2)}_H &=& \mu_0 {\bf \Phi_0 \Phi_0} + \mu_1 {\bf \Phi_1 
          \bar{\Phi}_1}
          + \mu_2 {\bf \Phi_2 \bar{\Phi}_2} + \mu_3 {\bf \Delta' 
\bar{\Delta}'}
          + \mu'_0 {\bf \Sigma_0\Sigma_0} + \mu'_1 {\bf \Sigma_1
          \bar{\Sigma}_1} + \mu'_2 {\bf \Sigma_2 \bar{\Sigma}_2} \nonumber\\
         & &  + \mu''_0 {\bf A_0 A_0} + \mu''_1 {\bf A_1 {\bar A}_1} + 
          \mu''_2 {\bf A_2 {\bar A}_2} + \mu''_3 {\bf A_3 {\bar A}_3} + 
          \mu'''_1 {\bf S_1 {\bar S}_1} + \mu'''_2 {\bf S_2 {\bar S}_2} + 
          \mu'''_3 {\bf S_3 {\bar S}_3} \\ \end{array} \eqno(2.11a)$$
\vspace{0.1in}
        $$\begin{array}{rcl}
        W^{(3)}_H &=& \lambda_0 {\bf \Phi_0 \Phi_0 \Phi_0} + \lambda_1 
          {\bf \Phi_1 \bar{\Phi}_1 \Phi_0} + \lambda_2 {\bf \Phi_2 
          \bar{\Phi}_2 \Phi_0} + \lambda_3 {\bf \Phi_1 \bar{\Phi}_2 
          \bar{\Phi}_2} + \lambda_4 {\bf \Phi_2 \Phi_2 \bar{\Phi}_1} 
          \nonumber\\
        & & + \rho_0 {\bf \Phi_0 \Phi_0 \Sigma_0} 
          + \rho_1 {\bf \Phi_1 \bar{\Phi}_1 \Sigma_0} 
          + \rho_2 {\bf \Phi_2 \bar{\Phi}_2 \Sigma_0} + \rho_3 {\bf \Phi_1
          \bar{\Phi}_2\bar{\Sigma}_2} + \rho_4 {\bf \Phi_2 \bar{\Phi}_1
          \Sigma_2} \\
        & & + \rho'_0 {\bf \Phi_0 \Phi_0 A_0} + \rho'_1 {\bf \Phi_1 
          \bar{\Phi}_1 A_0} + \rho'_2 {\bf \Phi_2 \bar{\Phi}_2 A_0}
          + \sigma_0 {\bf \Sigma_0 \Sigma_0 \Sigma_0} + \sigma_1 {\bf
          \Sigma_1 \bar{\Sigma}_1 \Sigma_0} \\
        & & + \sigma_2 {\bf \Sigma_2 \bar{\Sigma}_2 \Sigma_0} 
          + \sigma'_0 {\bf \Sigma_0 \Sigma_0 A_0} + 
          \sigma'_1 {\bf \Sigma_1 \bar{\Sigma}_1 A_0} + \sigma'_2 {\bf 
\Sigma_2
          \bar{\Sigma}_2 A_0} + \kappa_0 {\bf A_0 A_0 \Phi_0} \\
        \end{array}$$
        $$\begin{array}{rcl}
        & & + \kappa_1 {\bf A_1 \bar{A}_1 \Phi_0} 
          + \kappa_2 {\bf A_2 \bar{A}_2 \Phi_0} + \kappa_3 {\bf A_3 
          \bar{A}_3 \Phi_0} + \kappa'_0 {\bf A_0 A_0 \Sigma_0} + 
          \kappa'_1 {\bf A_1 \bar{A}_1 \Sigma_0} \nonumber\\
        & & + \kappa'_2 {\bf A_2 \bar{A}_2 \Sigma_0} 
          + \kappa'_3 {\bf A_3 \bar{A}_3 \Sigma_0} + \kappa'_4 {\bf A_3 A_3
          \Sigma_1} + \kappa'_5 {\bf \bar{A}_3 \bar{A}_3 \bar{\Sigma}_1} +
          \kappa''_0 {\bf A_0 A_0 A_0} \\
        & & + \kappa''_1 {\bf A_1 \bar{A}_1 A_0} 
          + \kappa''_2 {\bf A_2 \bar{A}_2 A_0} + \kappa''_3 {\bf A_3 
\bar{A}_3 
          A_0} + \eta_0 {\bf \Delta \bar{\Delta} \Phi_0} + \eta_1 {\bf 
\Delta 
          \bar{\Delta} A_0} \\
        & & + \eta'_0 {\bf \Delta' \bar{\Delta} \Phi_0} 
          + \eta'_1 {\bf \Delta' \bar{\Delta} A_0} + \tau_1 {\bf \Delta H_1
          \bar{\Phi}_1} + \tau_2 {\bf \bar{\Delta} \bar{H}_1 \Phi_1} +
          \tau_3 {\bf \Delta H_2 \bar{\Phi}_2} \\
        & & + \tau_4 {\bf \bar{\Delta} \bar{H}_2 \Phi_2} 
          + \delta_1 {\bf H_1 \bar{H}_1 \Sigma_0} + \delta_2 {\bf H_2 
          \bar{H}_2 \Sigma_0} + \delta_3 {\bf H_2 H_2 \bar{\Sigma}_1} + 
          \delta_4 {\bf \bar{H}_2 \bar{H}_2 \Sigma_1} \\
        & &  + \delta_5 {\bf \bar{H}_1 H_2 \Sigma_2} 
          + \delta_6 {\bf H_1 \bar{H}_2 \bar{\Sigma}_2} + \delta'_1 {\bf H_1
          \bar{H}_1 A_0} + \delta'_2 {\bf H_2 \bar{H}_2 A_0} + \gamma_0 
          {\bf A_0 A_0 A_0} \\
        & & + \gamma_1 {\bf A_0 A_1 \bar{A}_1} 
          + \gamma_2 {\bf A_0 A_2 \bar{A}_2} + \gamma_3 {\bf A_0 A_3 
          \bar{A}_3} + \varepsilon_1 {\bf S_1 S_2 \bar{S}_3} + 
\varepsilon_2 
          {\bf \bar{S}_1 \bar{S}_2 S_3} \\ \end{array} \eqno(2.11b)$$ 
All the $\mu$ parameters in (2.11a) are taken to be of order of the SUSY-GUT
scale.  We have not introduced corresponding superheavy mass terms for 
the 
${\bf H_1}, {\bf \bar{H}_1}, {\bf H_2}, {\bf \bar{H}_2}$ and ${\bf 
\Delta},\ 
{\bf \bar{\Delta}}$ superfields in order to keep components of them 
light. As
shown in Sect. III, we shall introduce a $Z_2$ discrete symmetery in 
order to 
place further restrictions on the terms which can appear in the 
superpotential.

The result of having so many Higgs and matter superfields in the model is 
to introduce a Landau singularity between the SUSY-GUT scale and the 
Planck scale.  But this should be the case, for the model is to be 
considered 
only an effective theory at best.  As pointed out earlier, the 
level-5 ${\bf \overline{126}}\  SO(10)$ multiplets do not arise naturally
in superstring models \cite{126} and must be treated as effective operators.
In order to see the origin of the singularity more quantitatively, we 
note 
the one-loop approximation to the renormalization group equation for the 
running $SO(10)$ gauge coupling is given by 
        $${dg_{10}\over{dt}} = {1\over{16\pi^2}}\left[N_{10} + 8 N_{45}
                + 12 N_{54} + 35 (N_{126} + N_{\overline{126}}) + 56 N_{210}
                + 2 (N_{16} + N_{\overline{16}}) - 24\right]g^3_{10} 
          \eqno(2.12)$$
With $N_{16} = 15,\ N_{\overline{16}} = 12,\ N_{10} = 4,\ N_{45} = 7, \ 
N_{54} = N_{210} = 5$, and $\ N_{126} = N_{\overline 126} = 2$, we find 
a Landau singularity arises at the energy scale 
        $$\mu = \mu_{10} \exp\left[{4\pi^2\over{285 
g^2_{10}(\mu_{10})}}\right]
                \eqno(2.13)$$
where $\mu_{10} = \Lambda_{SGUT} \simeq 2 \times 10^{16}$ GeV.  With a
gauge coupling of $g_{10}(\mu_{10}) = 0.67$, the singularity occurs within
a factor of 1.5 of the SUSY-GUT scale.  This value is close to the mass
scale assumed in \cite{an2} for the mass of conjugate fermions which pair
off and get superheavy and enter the Dimopoulos tree diagrams for the 
fermion mass matrix contributions.
The suggestion then is that the model, representing an effective theory,
perhaps arises from a superstring theory which becomes confining within
two orders of magnitude of the string scale.  The higher-dimensional Higgs
representations that appear phenomenologically in the model can then be 
regarded as composite states of the simpler confining theory 
holding above the singularity.  The possible existence of an infrared
fixed point structure at an energy scale beyond $10^{16}$ GeV has been
suggested and explored in models without grand unification by Lanzagorta 
and Ross \cite{Ross}.

\section{ONE PAIR OF LIGHT HIGGS DOUBLETS}
We now address the issue of how one can obtain just one pair of light Higgs
doublets, in order to preserve a satisfactory electroweak scale value for 
$\sin^2 \theta_W$ in evolution from the grand unification scale \cite{1pair}.
For this purpose we use the technique of Lee and Mohapatra \cite{LM} by 
constructing the doublet Higgsino mass matrix.  

As indicated in (2.7), Higgsino doublets arise from the ${\bf 10, 126,
\overline{126}}$ and ${\bf 210}$ representations of $SO(10)$.  If we drop the
tildes and order the bases for the columns and rows, respectively, 
according to 
        $$\begin{array}{rcl}
          B_u &=& \left\{\Phi_{1u},\ \bar{\Phi}_{1u},\ \Phi_{2u},\
          \bar{\Phi}_{2u},\ \Phi_{0u},\ \Delta'_u,\ \bar{\Delta}'_u,\ 
          \Delta_u,\ \bar{\Delta}_u,\ H_{1u},\ \bar{H}_{1u},\ H_{2u},\ 
          \bar{H}_{2u}\right\} \nonumber\\
          B_d &=& \left\{\bar{\Phi}_{1d},\ \Phi_{1d},\ \bar{\Phi}_{2d},\ 
          \Phi_{2d},\ \Phi_{0d},\ \bar{\Delta}'_d,\ \Delta'_d,\ 
          \bar{\Delta}_d,\ \Delta_d,\ \bar{H}_{1d},\ H_{1d},\ 
\bar{H}_{2d},\ 
          H_{2d}\right\} \\ \end{array} \eqno(3.1)$$
we find the $13 \times 13$ matrix separates into two block diagonal pieces,
the first $7 \times 7$ and the second $6 \times 6$.  Since the first 
submatrix
is full rank 7, the first 7 Higgsino doublets all become superheavy.  In 
order
for just one pair of Higgs doublets to remain light at the $\Lambda_{SGUT}$
scale, the second block diagonal matrix must be rank 5.  

To achieve that goal, we first introduce a $Z_2$ discrete symmetry 
\cite{discsym} whereby the following superfields are assigned the quantum 
number -1:
        $${\bf \bar{\Phi}_1,\ \Phi_2,\ \bar{\Phi}_2,\ \bar{\Sigma}_2,\ 
          \Delta,\ A_0,\ A_3,\ \bar{A}_3} \eqno(3.2)$$
while all other superfields are assigned the quantum number +1.  If we then
demand that the allowed $W^{(3)}_H$ cubic superpotential terms respect the
$Z_2$ symmetry, while the $W^{(2)}_H$ quadratic superpotential terms are
allowed to violate it softly \cite{soft} as considered, for example,  by 
Lee and Mohapatra \cite{LM}, Eq. (2.11a) remains unchanged and is 
repeated here for convenience 
        $$\begin{array}{rcl}
          W^{(2)}_H &=& \mu_0 {\bf \Phi_0 \Phi_0} + \mu_1 {\bf \Phi_1 
          \bar{\Phi}_1}
          + \mu_2 {\bf \Phi_2 \bar{\Phi}_2} + \mu_3 {\bf \Delta' 
\bar{\Delta}'}
          + \mu'_0 {\bf \Sigma_0\Sigma_0} + \mu'_1 {\bf \Sigma_1
          \bar{\Sigma}_1} + \mu'_2 {\bf \Sigma_2 \bar{\Sigma}_2} \nonumber\\
         & &  + \mu''_0 {\bf A_0 A_0} + \mu''_1 {\bf A_1 {\bar A}_1} + 
          \mu''_2 {\bf A_2 {\bar A}_2} + \mu''_3 {\bf A_3 {\bar A}_3} + 
          \mu'''_1 {\bf S_1 {\bar S}_1} + \mu'''_2 {\bf S_2 {\bar S}_2} + 
          \mu'''_3 {\bf S_3 {\bar S}_3} \\ \end{array} \eqno(3.3a)$$
while (2.11b) reduces to 
        $$\begin{array}{rcl}
        W^{(3)}_H &=& \lambda_0 {\bf \Phi_0 \Phi_0 \Phi_0} + 
          \lambda_2 {\bf \Phi_2 \bar{\Phi}_2
          \Phi_0} + \lambda_3 {\bf \Phi_1 \bar{\Phi}_2 \bar{\Phi}_2}
          + \rho_0 {\bf \Phi_0 \Phi_0 \Sigma_0} \nonumber\\
        & & + \rho_2 {\bf \Phi_2 \bar{\Phi}_2 \Sigma_0} + \rho_3 {\bf \Phi_1
          \bar{\Phi}_2\bar{\Sigma}_2} + \rho_4 {\bf \Phi_2 \bar{\Phi}_1
          \Sigma_2} + \rho'_1 {\bf \Phi_1 \bar{\Phi}_1 A_0} \\
        & & + \sigma_0 {\bf \Sigma_0 \Sigma_0 \Sigma_0} + \sigma_1 {\bf
          \Sigma_1 \bar{\Sigma}_1 \Sigma_0} + \sigma'_2 {\bf \Sigma_2
          \bar{\Sigma}_2 A_0} + \kappa_0 {\bf A_0 A_0 \Phi_0} \\
        & &  + \kappa_1 {\bf A_1 \bar{A}_1 \Phi_0} + \kappa_2 {\bf A_2 
          \bar{A}_2 \Phi_0} + \kappa_3 {\bf A_3 
          \bar{A}_3 \Phi_0} + \kappa'_0 {\bf A_0 A_0 \Sigma_0} \\
        & & + \kappa'_1 {\bf A_1 \bar{A}_1 \Sigma_0} + \kappa'_2 {\bf A_2 
          \bar{A}_2 \Sigma_0} 
          + \kappa'_3 {\bf A_3 \bar{A}_3 \Sigma_0} + \kappa'_4 {\bf A_3 A_3
          \Sigma_1} \\
        & & + \kappa'_5 {\bf \bar{A}_3 \bar{A}_3 \bar{\Sigma}_1} + \eta_1 
{\bf 
          \Delta \bar{\Delta} A_0} + \eta'_0 {\bf \Delta' 
          \bar{\Delta}' \Phi_0} + \tau_1 {\bf \Delta H_1 \bar{\Phi}_1} \\
        & & + \tau_2 {\bf \bar{\Delta} \bar{H}_1 \Phi_1} + \tau_3 {\bf 
          \Delta H_2 \bar{\Phi}_2} + \delta_1 {\bf H_1 
          \bar{H}_1 \Sigma_0} + \delta_2 {\bf H_2 
          \bar{H}_2 \Sigma_0} \\
        & &  + \delta_3 {\bf H_2 H_2 \bar{\Sigma}_1} + \delta_4 {\bf 
\bar{H}_2
          \bar{H}_2 \Sigma_1} + \delta_5 {\bf 
          \bar{H}_1 H_2 \Sigma_2} 
          + \varepsilon_1 {\bf S_1 S_2 \bar{S}_3} + \varepsilon_2 {\bf 
          \bar{S}_1 \bar{S}_2 S_3} \\ \end{array} \eqno(3.3b)$$ 
The two terms $\mu_1 {\bf \Phi_1 \bar{\Phi}_1}$ and $\mu'_2 {\bf \Sigma_2
\bar{\Sigma}_2}$ of (3.3a) violate $Z_2$ invariance and break the $Z_2$
symmetry softly.  

We shall assume that the VEV for ${\bf A_0}$, $<A_0>$, which helps to make
all colored Higgsino triplets superheavy does not contribute to the doublet
Higgsino mass matrix as explained in Sect. IV.  
The $6 \times 6$ doublet Higgsino submatrix then becomes 
        $${\cal M}_H = \left(\matrix{0 & 0 & 0 & 
{1\over{\surd{2}}}\tau_2(b_1 +
          c_1) & 0 & 0 \cr                       
          0 & 0 & {1\over{\surd{2}}}\tau_1 (\bar{b}_1 + \bar{c}_1) & 0 & 
          {1\over{\surd{2}}}\tau_3 (\bar{b}_2 + \bar{c}_2) & 0 \cr 
          0 & {1\over{\surd{2}}}\tau_2 (b_1 - c_1) & \delta_1 r_0 & 0 & 
          \delta_5 r_2 & 0 \cr
          {1\over{\surd{2}}}\tau_1 (\bar{b}_1 - \bar{c}_1) & 0 & 0 & 
          \delta_1 r_0 & 0 & 0 \cr
          0 & 0 & 0 & 0 & \delta_2 r_0 & \delta_4 r_1 \cr
          {1\over{\surd{2}}}\tau_3 (\bar{b}_2 - \bar{c}_2) & 0 & 0 & 
          \delta_5 r_2 & \delta_3 \bar{r}_1 & \delta_2 r_0 \cr}\right) 
          \eqno(3.4)$$
We have used the notation of (2.6) for the VEV's involved.
If we assume the chiral symmetry is broken so some $c_i \neq 0,$ and in 
particular that $\bar{c}_1 = - \bar{b}_1$ while $c_1 \neq \pm b_1$, the 
23 
element of the above
matrix vanishes, and we obtain a rank 5 matrix.  The massless Higgsino
doublet at the $\Lambda_{SGUT}$ scale is then given by 
        $$\tilde{H}_u = \alpha_{12}\tilde{\bar{\Delta}}_u + \alpha_{13}
                \tilde{H}_{1u} \eqno(3.5a)$$
while the other massless Higgsino doublet is obtained from the transpose of
${\cal M}_H$ and is found to be
        $$\tilde{H}_d = \alpha'_{11} \tilde{\bar{\Delta}}_d + 
\alpha'_{12} 
          \tilde{\Delta}_d + \alpha'_{14} \tilde{H}_{1d} + \alpha'_{15}
          \tilde{\bar{H}}_{2d} + \alpha'_{16} \tilde{H}_{2d} \eqno(3.5b)$$
The coefficients in the two expansions are related by 
        $$\alpha_{12} = - \sqrt{2}(\delta_1 r_0)/(\tau_2 (b_1 -
          c_1))\alpha_{13} \eqno(3.6a)$$
and by 
        $$\begin{array}{rcl}
          \alpha'_{11} &=& -{\sqrt{2}}/(\tau_2(b_1 + c_1))\left[\delta_5
                r_2 - \delta_1 r_0 {\tau_3 (\bar{b}_2 - \bar{c}_2)}/
                (\tau_1 (\bar{b}_1 - \bar{c}_1))\right]\alpha'_{16} 
\nonumber\\
          \alpha'_{12} &=& -{\sqrt{2}}/(\tau_3 (\bar{b}_2 + \bar{c}_2))
                \left[\delta_3 \bar{r}_1 - {\delta^2_2 r^2_0}/(\delta_4
                r_1)\right]\alpha'_{16} \\
          \alpha'_{14} &=& - {\tau_3 (\bar{b}_2 - \bar{c}_2)}/(\tau_1 
                (\bar{b}_1 - \bar{c}_1))\alpha'_{16} \\
          \alpha'_{15} &=& - {\delta_2 r_0}/(\delta_4 r_1)\alpha'_{16} \\
        \end{array} \eqno(3.6b)$$

Note that by our choice of chiral symmetry breaking, $\bar{c}_1 = - 
\bar{b}_1$,
for the VEV's of $\bar{\Phi}_1$, the corresponding Higgs doublet $H_u$ 
has 
components only in the $\bar{\Delta}_u$ and $H_{1u}$ directions, and can 
contribute only to the diagonal 33 and 
22 elements of the up quark and Dirac neutrino mass matrices in lowest-order
tree level as a result of the $U(1)_F$ charges.  On the other hand, the 
Higgs 
doublet $H_d$ has components in the $\bar{\Delta}_d,\ \Delta_d,\ H_{1d},
\ \bar{H}_{2d}$ and $H_{2d}$ directions, with lowest-order tree-level
contributions to all four (33, 23, 32 and 22) elements of the 
down quark and charged lepton mass matrices.  This helps to explain how 
it is
possible that the basis with up quark and Dirac neutrino mass matrices 
diagonal
can be selected as the preferred basis leading to simple $SO(10)$ mass
matrices.  For details see Ref. \cite{an1,an2}.

The other Higgsino doublets are superheavy and are general linear 
combinations
of all six basis vectors in the subspace.
        $$\eqalignno{
          \tilde{\cal H}_{iu} &= \alpha_{i1}\tilde{\Delta}_u + \alpha_{i2}
          \tilde{\bar{\Delta}}_u + \alpha_{i3}\tilde{H}_{1u} + \alpha_{i4}
          \tilde{\bar{H}}_{1u} + \alpha_{i5}\tilde{H}_{2u} + \alpha_{i6}
          \tilde{\bar{H}}_{2u},\qquad i = 2,3...6 &(3.7a)\cr
          \tilde{\cal H}_{id} &= \alpha'_{i1}\tilde{\bar{\Delta}}_d + 
          \alpha'_{i2}\tilde{\Delta}_d + \alpha'_{i3}\tilde{\bar{H}}_{1d} 
+ 
          \alpha'_{i4}\tilde{H}_{1d} + \alpha'_{i5}\tilde{\bar{H}}_{2d} + 
          \alpha'_{i6}\tilde{H}_{2d},\qquad i = 2,3...6 &(3.7b)\cr}$$
By inverting Eqs. (3.5) and (3.7), we obtain with suitable normalization
        $$\eqalignno{
          \tilde{\bar{\Delta}}_u &= \alpha^*_{12} \tilde{H}_u + \sum^6_{i=2}
                \alpha^*_{i2} \tilde{\cal H}_{iu} &(3.8a)\cr
          \tilde{H}_{1u} &= \alpha^*_{13} \tilde{H}_u + \sum^6_{i=2}
                \alpha^*_{i3} \tilde{\cal H}_{iu} &(3.8b)\cr
          \tilde{\Delta}_u,\ \tilde{\bar{H}}_{1u},\ \tilde{H}_{2u},\ 
                \tilde{\bar{H}}_{2u} &= \sum^6_{i=2} \alpha'^*_{ik}
                \tilde{\cal H}_{iu},\qquad k = 1,4,5,6 &(3.8c)\cr}$$
and 
        $$\eqalignno{
          \tilde{\bar{\Delta}}_d &= \alpha'^*_{11} \tilde{H}_d + \sum^6_{i=2}
                \alpha'^*_{i1} \tilde{\cal H}_{id} &(3.9a)\cr
          \tilde{\Delta}_d &= \alpha'^*_{12} \tilde{H}_d + \sum^6_{i=2}
                \alpha'^*_{i2} \tilde{\cal H}_{id} &(3.9b)\cr
          \tilde{\bar{H}}_{1d} &= \sum^6_{i=2} \alpha'^*_{i3}
                \tilde{\cal H}_{id} &(3.9c)\cr
          \tilde{H}_{1d} &= \alpha'^*_{14}\tilde{H}_d + \sum^6_{i=2}
                \alpha'^*_{i4} \tilde{\cal H}_{id} &(3.9d)\cr
          \tilde{\bar{H}}_{2d} &= \alpha'^*_{15}\tilde{H}_d + \sum^6_{i=2}
                \alpha'^*_{i5} \tilde{\cal H}_{id} &(3.9e)\cr
          \tilde{H}_{2d} &= \alpha'^*_{16}\tilde{H}_d + \sum^6_{i=2}
                \alpha'^*_{i6} \tilde{\cal H}_{id} &(3.9f)\cr}$$

The superheavy fields decouple at the $\Lambda_{SGUT}$ scale, and electroweak
VEV's are generated only by the light Higgs doublets as follows:
        $$\begin{array}{rclrcl}
          <\bar{\Delta}_u> &=& \alpha^*_{12}<H_u>, \qquad & <H_{1u}> 
                &=& \alpha^*_{13}<H_u> \nonumber\\
          <\bar{\Delta}_d> &=& \alpha'^*_{11}<H_d>, \qquad & <\Delta_d> &=&
                \alpha'^*_{12}<H_d> \\
          & & & <H_{1d}> &=& \alpha'^*_{14}<H_d> \\
          <\bar{H}_{2d}> &=& \alpha'^*_{15}<H_d>, \qquad & 
                <H_{2d}> &=& \alpha'^*_{16}<H_d> \\ \end{array} 
                \eqno(3.10)$$
We observe from the above that one pair of light Higgs doublets makes several
electroweak tree-level VEV contributions as found earlier in our $SO(10) 
\times U(1)_F$ model summarized in Sect. I.  Since the ${\bf 10}$ VEV's, 
$<H_{1u}>$ and $<H_{1d}>$, contributing to the 33 mass matrix elements 
are 
considerably larger than the ${\bf 10'}$ and ${\bf ^(\bar{126}^)}$ VEV's,
it is required from the above that $\tilde{H}_u$ and $\tilde{H}_d$ point 
mainly 
in the ${\bf 10}$ direction.

\section{SUPERHEAVY HIGGS TRIPLETS}
We now turn our attention to the Higgs doublet-triplet splitting problem.
The point is that unless all Higgs triplets get superheavy, too rapid proton
decay can take place by the exchange of a Higgsino color triplet leading to
a dimension-5 contribution or by the exchange of a Higgs color triplet
leading to a dimension-6 contribution to proton decay \cite{colHiggs}.  
This problem can be alleviated through the Dimopoulos - Wilczek type 
mechanism \cite{DW}.

The Higgsino triplets appear in the representations singled out in (2.9).
We thus choose to order the bases for the triplet Higgsino mass matrix as 
follows where we again have dropped tildes:
        $$\begin{array}{rcl}
          B_u &=& \left\{\Phi_{1t},\ \bar{\Phi}_{1t},\ \Phi_{2t},
          \ \bar{\Phi}_{2t},\ \Phi_{0t},\ \Delta'^{(1,1,6)}_t,\ \bar{\Delta}
          ^{'(1,1,6)}_t,\ \bar{\Delta}'^{(1,3,10)}_t,\right. \nonumber\\
        & & \qquad \left.{\Delta^{(1,1,6)}_t,\ \bar{\Delta}^{(1,1,6)}_t,
          \ \bar{\Delta}^{(1,3,10)},\ H_{1t},\ \bar{H}_{1t},\ H_{2t},
          \ \bar{H}_{2t}}\right\} \\ \end{array} \eqno(4.1a)$$
and
        $$\begin{array}{rcl}
          B_d &=& \left\{\bar{\Phi}_{1\bar{t}},\ \Phi_{1\bar{t}},\ 
\bar{\Phi}_
          {2\bar{t}},\ \Phi_{2\bar{t}},\ \Phi_{0\bar{t}},\ \bar{\Delta}
          ^{'(1,1,6)}_{\bar{t}},\ \Delta'^{(1,1,6)}_{\bar{t}},
          \ \Delta'^{(1,3,\overline{10})}_{\bar{t}},\right. \nonumber\\
        & & \qquad \left.{\bar{\Delta}^{(1,1,6)}_{\bar{t}},\ 
\Delta^{(1,1,6)}_
          {\bar{t}},\ \Delta^{(1,3,\overline{10})}_{\bar{t}},\ 
\bar{H}_{1\bar{
          t}},\ H_{1\bar{t}},\ \bar{H}_{2\bar{t}},\ H_{2\bar{t}}}\right\} 
\\ 
          \end{array} \eqno(4.1b)$$

We now assume that the Dimopoulos - Wilczek type mechanism operates, so the
VEV for ${\bf A_0}$ takes the form \\[-0.2in]
        $$<A_0> = diag(0,\ 0,\ a,\ a,\ a) \otimes \epsilon = p_0 a_{1,1,15}
          \eqno(4.2)$$
where $\epsilon$ is the $2 \times 2$ antisymmetric matrix.  This then 
contributes to the colored triplet Higgsino mass matrix but not to the 
doublet
Higgsino mass matrix given in (3.4).  If such is the case, the colored 
triplet
Higgsino mass matrix splits into two block diagonal submatrices of 
dimensions 
$8 \times 8$ and $7 \times 7$ in terms of the bases given above.  The 
first is
trivially full rank, while the second assumes the following form:\\[0.1in]
${\cal M}_{H'} =$ \hfill (4.3)
        $$\left(\matrix{\eta_1 p_0 & 0 & 0 & 0 & {1\over{
          \surd{2}}}\tau_2 (a_1 + b_1) & 0 & 0 \cr 
          0 & \eta_1 p_0 & 0 & {1\over{\surd{2}}}\tau_1 (\bar{a}_1 + 
\bar{b}_1)
          & 0 & {1\over{\surd{2}}}\tau_3 (\bar{a}_2 + \bar{b}_2) & 0 \cr 
          0 & 0 & \eta_1 p_0 & \tau_1 \bar{c}_1 & 0 & \tau_3 \bar{c}_2 & 
0 \cr
          0 & {1\over{\surd{2}}}\tau_2 (a_1 - b_1) & \tau_2 c_1 & 
          \delta_1 r_0 & 0 & \delta_5 r_2 & 0 \cr
          {1\over{\surd{2}}}\tau_1 (\bar{a}_1 - \bar{b}_1) & 0 & 0 & 0 & 
          \delta_1 r_0 & 0 & 0 \cr
          0 & 0 & 0 & 0 & 0 & \delta_2 r_0 & \delta_4 r_1 \cr
          {1\over{\surd{2}}}\tau_3 (\bar{a}_2 - \bar{b}_2) & 0 & 0 & 0 & 
          \delta_5 r_2 & \delta_3 \bar{r}_1 & \delta_2 r_0 \cr}\right)$$ 
\vspace{0.1in}
By inspection the above matrix is also full rank, so all color triplet 
Higgsinos become superheavy.   The important point is that 
$<A_0> = p_0 a_{1,1,15}$ does not contribute a mass
term to the ${\bf ^{(}\overline{126}^{)}}\ (2,2,15)$ Higgsino doublets, 
since 
the $SU(4)$ Clebsch-Gordan coefficient yielding an antisymmetric ${\bf 45}$
representation vanishes \cite{Mof}.  Thus splitting of one pair of 
doublet and
triplet Higgsinos may be achieved through a Dimopoulos - Wilczek type 
mechanism, though the special conditions required in the work of Babu and 
Barr
\cite{colHiggs} make this somewhat problematic.

\section{GUT SCALE CONDITIONS FOR WEAK SCALE SUPERSYMMETRY}
We now turn our attention to the subject of weak scale supersymmetry and the
conditions which must obtain for the supersymmetry to remain unbroken at 
the $\Lambda_{SGUT}$ scale.  We are referring to the conditions which 
preserve
some F-flat and D-flat directions for which the minimum $V = 0$ of the 
scalar 
potential is maintained \cite{SUSY}.  This requires that 
        $$V(\{\phi_i\}) = \sum_i |F_i|^2 + {1\over{2}}\sum_a |D^a|^2 
                \eqno(5.1a)$$
vanishes for the directions singled out by the VEV's of the scalar fields.
The sum goes over all fields present in the Higgs superpotential, where
        $$F_i = {\partial W\over{\partial \phi_i}}, \qquad 
          D^a = - g \phi^*_i T^a_{ij} \phi_j \eqno(5.1b)$$
For the purposes of this Section, we have ignored any explicit soft 
supersymmetry-breaking terms.

The F-terms appearing in (5.1) then involve the following derivatives as
indicated by an obvious shorthand notation:
        $$\begin{array}{rll}
          &F_{\Phi_0},\ F_{\Phi_i},\ F_{\bar{\Phi}_i},\qquad &i = 1,2 
                \nonumber\\
          &F_{\Delta'},\ F_{\bar{\Delta}'}, \cr
          &F_{\Sigma_0},\ F_{\Sigma_i},\ F_{\bar{\Sigma}_i},\qquad &i = 
1,2 \\
          &F_{A_0},\ F_{A_i},\ F_{\bar{A}_i},\qquad &i = 1,2,3 \\
          &F_{S_0},\ F_{S_i},\ F_{\bar{S}_i},\qquad &i = 1,2,3 \\
          \end{array} \eqno(5.2)$$
We have written down the F-flat conditions in terms of the VEV's 
appearing in 
(2.6) and kept only those terms whose $\Lambda_{SGUT}$ VEV's are 
non-vanishing
\cite{Lee}.  For $\{F_{\Phi_i},\ F_{\bar{\Phi}_i}\}$ and $\{F_{A_i},
\ F_{\bar{A}_i}\}$ for each i there are three and two conditions, 
respectively,
since the coefficient for each possible VEV direction must vanish.  For 
$F_{\Sigma_i}$ and $F_{\bar{\Sigma}_i}$, two conditions
also arise, for the contributions point not only in the $\sigma_{1,1,1}$
direction, but also in the $s_{1,1,1}$ direction.  Note that the conditions
allow all the masses present in (3.3a) to be superheavy, while the VEV's
in (2.6) are also near the $\Lambda_{SGUT}$ scale.  No F-flat directions are
lifted in so doing.  Nor are any Goldstone bosons introduced by the $SO(10)$
symmetry breaking.

In order for $p_0 \ne 0,\ q_0 = 0$ to be satisfied so all colored Higgs 
triplets are superheavy while one pair of Higgs doublets can remain 
light, 
we must set $c_0 = 0$.  Consistency of the remaining conditions is easily 
maintained by setting $\lambda_3 = 0$.  Some additional simple relations 
that 
follow for self-consistency are
        $$\eqalignno{
          {{a_1}\over{a_2}} =&- {3\over{2}}{{b_1}\over{b_2}} = 6 {{c_1}\over{
                c_2}} &(5.3a)\cr
          {{\bar{a}_1}\over{\bar{a}_2}} =&- 
{3\over{2}}{{\bar{b}_1}\over{{\bar{
                b}_2}}} = 6 {{\bar{c}_1}\over{\bar{c}_2}} &(5.3b)\cr
          \rho_3 \bar{r}_2 (\bar{a}_2 / \bar{a}_1) =&\rho_4 r_2 (a_2 /a_1)
                &(5.3c)\cr
          {p_3} /{\bar{p}_3} =&{q_3} /{\bar{q}_3} &(5.3d)\cr
          \kappa'_4 r_1 p^2_3 =&\kappa'_5 \bar{r}_1 {\bar{p}^2_3} &(5.3e)
                \cr
          \mu_3 =&-{1\over{10}}\eta'_0 \left[{1\over{\surd{6}}}a_0 
                + {1\over{\surd{2}}}b_0\right] &(5.3f)\cr 
          \mu'_1 =&- {1\over{2\sqrt{15}}} \kappa'_4 (3p_3^2 - 2q^2_3)/
                {\bar{r}_1} &(5.3g)\cr
          \mu'_2 =&{1\over{\sqrt{15}}}\rho_4 (a_1 \bar{a}_1 + b_1 \bar{b}_1
                + c_1 \bar{c}_1) (c_2 /c_1 \bar{r}_2) &(5.3h)\cr
          \kappa'_1 / \kappa_1 = \kappa'_2 / \kappa_2 =&\kappa'_3 / \kappa_3
                + 2 (\kappa'_4 r_1 p_3 )/ (\kappa_3 r_0 \bar{p}_3) = 
                \left[\sqrt{2\over{5}}a_0 - 
{2\surd{2}\over{\sqrt{15}}}b_0 
                \right]{1\over{r_0}} &(5.3i)\cr
          \mu''_0 =&-{2\over{3\surd{2}}}\kappa_0 b_0 - {1\over{\surd{15}}}
                \kappa'_0 r_0  &(5.3j)\cr
          \mu''_1 / \kappa_1 = \mu''_2 / \kappa_2 =&\mu''_3 / \kappa_3 
                = - {2\over{5}}\left[{1\over{\surd{2}}}b_0 + 
{1\over{\surd{6}}}
                a_0 \right] &(5.3k)\cr
          \mu'''_1 t_1 \bar{t}_1 =&\mu'''_2 t_2 \bar{t}_2 = \mu'''_3 t_3
                \bar{t}_3  &(5.3l)\cr}$$

The additional restriction that $\bar{b}_1 = - \bar{c}_1$, needed to 
ensure 
that only one pair of Higgs doublets remains light, further implies that 
$4\bar{b}_2 = \bar{c}_2$.  No restrictions are found on $p_i,\ \bar{p}_i,
\ q_i,\ \bar{q}_i$ for $i = 1,2$ which appear in the VEV's of the ${\bf 
45}$'s 
needed to break the $SO(10)$ symmetry down to the SM at the 
$\Lambda_{SGUT}$ 
scale.  Several special cases of interest for the ${\bf 45}$ VEV's in 
addition
to that employed for $A_0$ in (4.2) are the following:
        $$\begin{array}{rll}
          <A_{45_d}> =&diag(q,\ q,\ 0,\ 0,\ 0) \otimes \epsilon &\sim
                qa_{1,3,1} \nonumber\\
          <A_{45_X}> =&diag(u,\ u,\ u,\ u,\ u) \otimes \epsilon &\sim
                \left(\sqrt{{3\over{5}}} a_{1,1,15} + \sqrt{{2\over{5}}}
                a_{1,3,1}\right)u \\
          <A_{45_Y}> =&diag(3u,\ 3u,\ -2u,\ -2u,\ -2u) \otimes \epsilon &\sim
                \left(\sqrt{{2\over{5}}} a_{1,1,15} - \sqrt{{3\over{5}}}
                a_{1,3,1}\right)u \\
          <A_{45_Z}> =&diag(3u,\ 3u,\ 2u,\ 2u,\ 2u) \otimes \epsilon &\sim
                \left(\sqrt{{2\over{5}}} a_{1,1,15} + \sqrt{{3\over{5}}}
                a_{1,3,1}\right)u \\
        \end{array} \eqno(5.4)$$
In \cite{an2} we have chosen the VEV's in the $A_{45_X}$ and $A_{45_Z}$ 
directions to be non-vanishing, so the $SO(10)$ symmetry is broken 
directly to 
the SM: $SO(10) \rightarrow SU(3)_c \times SU(2)_L \times U(1)_Y$.  While 
such 
VEV's appear to be allowed by our analysis, unfortunately they are not 
uniquely singled out.

If we gauge the $U(1)_F$ family symmetry, D-terms can arise from the 
spontaneous breaking of the $U(1)_F$ and $SO(10)$ at the SUSY-GUT scale
and involve only $D_F$ and $D_X$, if $SO(10) \times U(1)_F$ breaks directly
to the SM as we have assumed in \cite{an2}.  These terms will vanish in 
the 
limit that the soft supersymmetry breaking terms are neglected, as the 
VEV's for the conjugate fields $\phi_i$ and $\bar{\phi}_i$ which break the
$U(1)$ symmetries become equal.  We shall address the soft 
supersymmetry-breaking in the next Section.

\section{SOFT SUSY-BREAKING CONTRIBUTIONS}
Here we present the supersymmetric part of the scalar potential which applies
when the supersymmetry is softly broken:    
        $$V(\{\phi_i\}) = \sum_i |F_i|^2 + {1\over{2}}\sum_a |D^a|^2 
                + V_{\rm soft}  \eqno(6.1a)$$
where 
        $$F_i = {\partial W\over{\partial \phi_i}}, \qquad 
          D^a = - g \phi^*_i T^a_{ij} \phi_j, \eqno(6.1b)$$

The soft SUSY-breaking part of the scalar potential, so far as the Higgs mass
terms are concerned, is given by 
        $$\begin{array}{rcl}
          V_{soft} &=& m^2_0 |\Phi_0|^2 + m^2_1 |\Phi_1|^2 + \bar{m}^2_1
                |\bar{\Phi}_1|^2 + m^2_2 |\Phi_2|^2 + \bar{m}^2_2
                |\bar{\Phi}_2|^2 + m^2_3 |\Delta'|^2 + \bar{m}^2_3
                |\bar{\Delta}'|^2 \nonumber\\
                & & \quad + m'^2_0 |\Sigma_0|^2 + m'^2_1 |\Sigma_1|^2
                + \bar{m}'^2_1 |\bar{\Sigma}_1|^2 + m'^2_2 |\Sigma_2|^2 + 
                \bar{m}'^2_2 |\bar{\Sigma}_2|^2 
                + m''^2_0 |A_0|^2 \\
                & & \quad + m''^2_1 |A_1|^2 
                + \bar{m}''^2_1 |\bar{A}_1|^2 + m''^2_2 |A_2|^2 + 
                \bar{m}''^2_2 |\bar{A}_2|^2 + m''^2_3 |A_3|^2 + 
                \bar{m}''^2_3 |\bar{A}_3|^2 \\
                & & \quad + m'''^2_1 |S_1|^2 + 
                \bar{m}'''^2_1 |\bar{S}_1|^2 + m'''^2_2 |S_2|^2 + 
                \bar{m}'''^2_2 |\bar{S}_2|^2 + m'''^2_3 |S_3|^2 + 
                \bar{m}'''^2_3 |\bar{S}_3|^2 \\
                & & \quad + m^2_u |H_u|^2 + m^2_d |H_d|^2
                + m^2_{ud} (\epsilon_{ij}H^i_u H^j_d + h.c.)\\
          \end{array} \eqno(6.2)$$
The D-terms include contributions from the broken $U(1)_F$ and $U(1)_X$,
as well as the $SU(2)_L$ and $U(1)_Y$, which are given by
        $$\begin{array}{rcl}
          V_D &=& {1\over{2}}g^2_F \left[2(|S_1|^2 - |\bar{S}_1|^2) + 
                6.5(|S_2|^2 - |\bar{S}_2|^2) + 8.5(|S_3|^2 - |\bar{S}_3|^2)
                \right. \nonumber\\
                & &\quad + 3.5(|A_1|^2 - |\bar{A}_1|^2) + 0.5(|A_2|^2 - 
                |\bar{A}_2|^2)
                - 16(|\Sigma_1|^2 - |\bar{\Sigma}_1|^2) \\
                & & \quad - 10(|\Sigma_2|^2 - 
                |\bar{\Sigma}_2|^2) - 20(|\Phi_1|^2 - |\bar{\Phi}_1|^2) 
                - 10(|\Phi_2|^2 - |\bar{\Phi}_2|^2) \\
                & & \quad + 22(|\Delta'|^2 - |\bar{\Delta}'|^2) 
                - 2(|\Delta|^2 - |\bar{\Delta}|^2) 
               - 18(|H_1|^2 - |\bar{H}_1|^2) \\
                & & \quad \left.- 8(|H_2|^2 - 
                |\bar{H}_2|^2)
                - 8|\tilde{\psi}_1|^2 - |\tilde{\psi}_2|^2 
                + 9|\tilde{\psi}_3|^2 - 12(|\tilde{n}|^2 + |\tilde{n}^c|^2)
                \right]^2 \\
              & & \quad + {1\over{2}}g^2_X \left[-10(|\Delta'|^2 - 
                |\bar{\Delta}'|^2) - 2(|\Delta|^2 - |\bar{\Delta}|^2) + 
                2(|H_1|^2 - |\bar{H}_1|^2) \right. \\
                & & \quad \left. + 2(|H_2|^2 - |\bar{H}_2|^2) + 
...\right]^2 \\
              & & \quad + {1\over{8}}g^2 \left[|H_u|^4 + |H_d|^4 - 2|H_u|^2
                |H_d|^2 + 4|H^{\dagger}_uH_d|^2 \right] \\
              & & \quad + {1\over{8}}g'^2 \left[|H_u|^4 + |H_d|^4 
                - 2|H_u|^2|H_d|^2 \right] \\
          \end{array} \eqno(6.3)$$
Once the soft SUSY-breaking masses are allowed to become nonuniversal, 
sizable D-term contributions to the scalar potential can result.
The F-terms can be found by differentiating the last few terms in
(3.3b) which are linear in one superheavy field with respect to that 
field.   
We find 
        $$\begin{array}{rcl}
          V_F &=& |\tau_1 \Delta H_1|^2 + |\tau_2 \bar{\Delta}\bar{H}_1|^2
                + |\delta_1 H_1 \bar{H}_1 + \delta_2 H_2 \bar{H}_2|^2 
                \nonumber\\
              & &\qquad + |\delta_3 H_2 H_2|^2 + |\delta_4 
                \bar{H}_2\bar{H}_2|^2 + |\delta_5 \bar{H}_1 H_2|^2 + 
|\eta_1 
                \Delta \bar{\Delta}|^2 \\
          \end{array} \eqno(6.4)$$
Upon minimizing the full scalar potential, one finds the VEV's generated 
for 
the scalar fields and their conjugates become unequal provided some 
$m^2$'s 
are driven negative as shown in \cite{DKM}.  Supersymmetry is broken along
a nearly D-flat direction with $|m| = O(1\ {\rm TeV})$.  

By making use of (3.8)
and (3.9) to replace the original Higgs doublets by the pair $H_u$ and
$H_d$ which remains light down to the electroweak scale and integrating out
the fields which become superheavy, we find 
the scalar potential for the Higgs sector can be written as
        $$\begin{array}{rcl}
          V(Higgs) &=& m^2_u (H^{\dagger}_u H_u) + m^2_d (H^{\dagger}_d H_d)
                   + m^2_{12} (\varepsilon_{ij} H^i_u H^j_d + h.c.) 
\nonumber\\
                & & + {1\over{8}}(g^2 + g'^2) \left[(H^{\dagger}_u H_u)^2 
+ 
                   (H^{\dagger}_d H_d)^2 - 2 (H^{\dagger}_u 
H_u)(H^{\dagger}_d
                   H_d)\right] \cr
                & & + ({1\over{2}}g^2 + g'^2)|H^{\dagger}_u H_d|^2 \cr
          \end{array} \eqno(6.5)$$
Despite the apparent non-minimal nature of our model at the SUSY-GUT 
scale 
due to the presence of many Higgs contributions, since only one pair of 
Higgs 
doublets survives at the electroweak scale under the assumptions developed
in Sect. III, the scalar potential at that scale
is similar to that of the minimal supersymmetric standard model.  Thus the
good result for $\sin^2 \theta_W$ achieved in the MSSM is maintained, and
the evolution of all the gauge and Yukawa couplings from $\Lambda_{SGUT}$ 
is unaltered.

In integrating out the superheavy fields, one also finds nonuniversal 
corrections to the squark and slepton fields given by 
        $$\Delta m^2_a = Q_{Fa}D_F + Q_{Xa}D_X \eqno(6.6)$$
in the notation of Kolda and Martin \cite{DKM}, where the $Q$'s are the 
$U(1)_F$ and $U(1)_X$ charges and the $D$'s are parameters which summarize
the symmetry-breaking process at the SUSY-GUT scale.  The main point we
wish to make here is that the first and third family squark and slepton 
masses will
be split further away from their universal values than the second family,
due to their larger $U(1)_F$ charges.  Recall $Q_F = -8,\ -1,\ 9$ for
the first, second and third family, respectively.
Which family emerges with the smallest
mass depends on the sign of $D_F$.  In any case, the splitting will be 
limited 
by the present experimental constraints on flavor-changing neutral 
currents.\\  
\section{YUKAWA SUPERPOTENTIAL}
The superpotential for the Yukawa interactions can be simply constructed 
from the superfields introduced earlier, where every term remains invariant
under the $U(1)_F$ and $Z_2$ symmetries.  For this purpose we assign a 
$Z_2$ charge of +1 to each of the matter superfields ${\bf \Psi_i,\ F}^{(k)}$
and ${\bf \bar{F}}^{(k)}$.  We then find for the Yukawa superpotential
        $$\begin{array}{rl}
          W_Y =&g_{10} {\bf \Psi_3} {\bf \Psi_3} {\bf H_1} 
                + g_{10'}\left\{\left[{\bf \Psi_2} {\bf \Psi_3} + {\bf F}^{
                (-4.5)}{\bf F}^{(12.5)} + {\bf F}^{(4)}{\bf F}^{(4)} 
                + {\bf \bar{F}}^{(0.5)} {\bf \bar{F}}^{7.5)}\right]{\bf H_2}
                \right. \nonumber\\
          &\qquad\qquad\qquad\qquad\qquad + \left.\left[{\bf F}^{(-0.5)} 
                {\bf F}^{(-7.5)} + {\bf 
                \bar{F}}^{(4.5)} {\bf \bar{F}}^{(-12.5)} + {\bf 
\bar{F}}^{(-4)}
                {\bf \bar{F}}^{(-4)}\right]{\bf \bar{H}_2} \right\} \\
          +& g_{126}\left[{\bf \Psi_2}{\bf \Psi_2} + {\bf F}^{
                (-6)}{\bf F}^{(4)} + {\bf F}^{(4.5)}{\bf F}^{(-6.5)}\right]
                {\bf \bar{\Delta}} + g_{126'}\left[{\bf F}^{(11)}
                {\bf F}^{(11)}{\bf \bar{\Delta}'} 
                + {\bf \bar{F}}^{(-11)}{\bf \bar{F}}^{(-11)}{\bf
                \Delta'}\right] \\
          +& g'_{45}\left\{\left[{\bf \Psi_1}{\bf \bar{F}}^{(4.5)}
                + {\bf \Psi_3}{\bf \bar{F}}^{(-12.5)} + {\bf F}^{(1)}
                {\bf \bar{F}}^{(-4.5)}\right]{\bf A_1} + \left[{\bf \Psi_2}
                {\bf \bar{F}}^{(4.5)} + {\bf F}^{(4.5)}{\bf \bar{F}}^{(-1)}
                \right]{\bf \bar{A}_1} \right\} \\
          +& g''_{45}\left\{\left[{\bf \Psi_2}{\bf \bar{F}}^{(0.5)}
                + {\bf \Psi_1}{\bf \bar{F}}^{(7.5)} + {\bf F}^{(1)}{\bf \bar{
                F}}^{(-1.5)} + {\bf F}^{(4)}{\bf \bar{F}}^{(-4.5)}
                + {\bf F}^{(1.5)}{\bf \bar{F}}^{(-2)} + {\bf 
F}^{(-6.5)}{\bf 
                \bar{F}}^{(6)}\right]{\bf A_2}\right. \\
          &\qquad\qquad + \left. \left[{\bf F}^{(2)}{\bf 
                \bar{F}}^{(-1.5)} + {\bf F}^{(4.5)}{\bf \bar{F}}^{(-4)}
                + {\bf F}^{(1.5)}{\bf \bar{F}}^{(-1)} + {\bf 
F}^{(-6)}{\bf 
                \bar{F}}^{(6.5)}\right]{\bf \bar{A}_2}\right\} \\
          +& g'_{1}\left\{ \left[{\bf \Psi_3 \bar{F}}^{(-11)} 
                + {\bf \Psi_2 \bar{F}}^{(-1)} + {\bf \Psi_1 
\bar{F}}^{(6)} 
                + {\bf F}^{(-0.5)}{\bf \bar{F}}^{(-1.5)} + {\bf 
F}^{(2)}{\bf 
                \bar{F}}^{(-4)} + {\bf F}^{(-6.5)}{\bf \bar{F}}^{(4.5)}
                \right]{\bf S_1}\right. \\ 
          &\qquad\qquad\qquad + \left. \left[{\bf F}^{(4)}{\bf 
\bar{F}}^{(-2)}
                + {\bf F}^{(-4.5)}{\bf \bar{F}}^{(6.5)} + {\bf F}^{(1.5)}
                {\bf \bar{F}}^{(0.5)}\right]{\bf \bar{S}_1}\right\} \\
          +& g''_{1}\left\{\left[{\bf F}^{(4.5)}{\bf \bar{F}}^{(-11)}
                + {\bf F}^{(-4.5)}{\bf \bar{F}}^{(-2)}\right]{\bf S_2}
                + \left[{\bf \Psi_2 \bar{F}}^{(7.5)} + {\bf F}
                ^{(2)}{\bf \bar{F}}^{(4.5)} + {\bf F}^{(11)}{\bf \bar{F}}^{
                (-4.5)}\right]{\bf \bar{S}_2}\right\} \\
           \end{array}$$
          $$\begin{array}{rl}
          +& g'''_{1}\left\{\left[{\bf F}^{(4)}{\bf \bar{F}}^{(-12.5)}
                + {\bf F}^{(-4.5)}{\bf \bar{F}}^{(-4)} + {\bf 
F}^{(-7.5)}{\bf 
                \bar{F}}^{(-1)} + {\bf F}^{(-6.5)}{\bf \bar{F}}^{(-2)}\right]
                {\bf S_3}\right. \\ 
          &\qquad\qquad\qquad + \left. \left[{\bf F}^{(1)}{\bf 
\bar{F}}^{(7.5)}
                + {\bf F}^{(2)}{\bf \bar{F}}^{(6.5)} + {\bf F}^{(4)}{\bf 
\bar{
                F}}^{(4.5)} + {\bf F}^{(12.5)}{\bf \bar{F}}^{(-4)}\right]
                {\bf \bar{S}_3}\right\} \\ 
          +& g_{210}\left[{\bf F}^{(-0.5)}{\bf \bar{F}}^{(0.5)} + {\bf F}^{
                (1)}{\bf \bar{F}}^{(-1)} + {\bf F}^{(2)}{\bf \bar{F}}^{(-2)}
                + {\bf F}^{(4)}{\bf \bar{F}}^{(-4)} \right. \nonumber\\
          &\qquad\qquad\qquad + {\bf F}^{(4.5)}{\bf \bar{F}}^{(-4.5)} 
                + {\bf F}^{(-4.5)}{\bf \bar{F}}^{(4.5)}
                + {\bf F}^{(-7.5)}{\bf \bar{F}}^{(7.5)} + {\bf 
F}^{(11)}{\bf 
                \bar{F}}^{(-11)} \\
          &\qquad\qquad\qquad + \left. {\bf F}^{(12.5)}{\bf 
\bar{F}}^{(-12.5)}
                + {\bf F}^{(1.5)}{\bf \bar{F}}^{(-1.5)} + {\bf 
F}^{(-6)}{\bf 
                \bar{F}}^{(6)} + {\bf F}^{(-6.5)}{\bf \bar{F}}^{(6.5)}
                \right]{\bf \Phi_0} \\
          +& g'_{210}\left[{\bf F}^{(12.5)}{\bf \bar{F}}^{(7.5)}\right]
                {\bf \Phi_1} \\
        \end{array} \eqno(7.1)$$
where we have assumed the Yukawa couplings are real.  All but the last three
terms involving ${\bf ^{(}\bar{S}^{)}_3,\ \Phi_0}$ and ${\bf \Phi_1}$ have
previously appeared in the $SO(10) \times U(1)_F$ model constructed earlier
in \cite{an2}.  These new terms can alter the numerical results previously
obtained in that reference if their corresponding Yukawa couplings do not
vanish; their effects will be discussed elsewhere.
\section{SUMMARY}
In this paper, as the third stage of an extended program, the author has 
explored many of the ingredients necessary to construct a supersymmetric 
grand unified model in the $SO(10) \times U(1)_F$ framework, based on the 
results obtained earlier with a bottom-up approach carried out in 
collaboration with S. Nandi.  In that earlier work, supersymmetry simply
controlled the running of the Yukawa couplings and enabled us to restrict
our attention to Dimopoulos tree diagrams to evaluate various mass matrix
elements.  Here we introduce complete supermultiplets, a superpotential
and soft-breaking terms in order to study more thoroughly the 
consequences of
such a SUSY-GUT model.  We have also pointed out some of the shortcomings
in our analysis, since the complexity of the model has not enabled us to 
test whether the desired symmetry-breaking directions can be achieved 
given all the conditions imposed.  

We started with the ${\bf 16}$ and ${\bf \overline{16}}$ 
fermion and ${\bf 1}$, ${\bf 10}$, ${\bf 45}$ and ${\bf \overline{126}}$
Higgs multiplets and their
associated $U(1)_F$ family charges required in \cite{an2} for the 
$SO(10) \times U(1)_F$
model construction of the quark and lepton mass matrices.  We extend these
same assignments to $SO(10)$ supermultiplets and add $U(1)_F$-conjugate Higgs
supermultiplets to make the $[SO(10)]^2 \times U(1)_F$ triangle anomaly
vanish.  The $[U(1)_F]^3$ triangle anomaly will also vanish, so the model 
is 
anomaly-free with the addition of just one pair of $SO(10)$ singlet 
supermultiplets,
both with $U(1)_F$ charge -12. Since these supermultiplets correspond to 
a sterile neutrino, a conjugate sterile neutrino and their scalar 
neutrino 
partners, but with the same $U(1)_F$ charges, they do not pair off and 
get 
superheavy. 

To this set of supermultiplets derived from the Yukawa sector of the model
must be added additional pairs of $U(1)_F$-conjugate Higgs supermultiplets
belonging to ${\bf 54}$ and ${\bf 210}$ representations for the Higgs sector
of the superpotential.  These are needed in order to generate appropriate 
higgsino mass matrices and to ensure that some 
F-flat direction exists after the breaking of the GUT symmetry, so that 
the supersymmetry remains unbroken at the $\Lambda_{SGUT}$ scale with its
breaking occurring in the visible sector only near the electroweak scale.

The large multiplicity of superfields introduced results in the development
of a Landau singularity within a factor of 1.5 of $\Lambda_{SGUT}$ when the
$SO(10)$ gauge coupling is evolved beyond the SUSY-GUT scale toward the 
Planck
scale.  We have argued that this should occur, for the model is an effective
theory at best since the higher level $SO(10)$ supermultiplets do not 
arise naturally in superstring models, for example.  The appearance of 
the 
Landau singularity suggests that the true theory near the Planck scale
becomes confining when evolved downward through two orders of magnitude
with the higher-dimensional Higgs representations emerging as composite
states of that theory.  

By the introduction of a $Z_2$ discrete symmetry and the judicious choice 
of chiral symmetry breaking, we find that the it can be arranged that only
one pair of higgsino (Higgs) doublets remains light at the electroweak scale;
on the other hand, all higgsino triplets become superheavy.  Moreover,
the electroweak VEV's generated by the light pair of Higgs doublets make
lowest-order tree-level contributions only to the diagonal 22 and 33 
elements 
of the up quark and Dirac
neutrino mass matrices, while all four elements in the 2-3 sector of the 
down quark and charged lepton mass matrices receive such tree-level 
contributions.
This is in agreement with the phenomenological results obtained earlier 
in Ref. \cite{an2}.  

By the addition of soft SUSY-breaking terms to the scalar part of the Higgs
potential, nonuniversal corrections to the masses of the squark and 
slepton 
fields can be generated which involve the $U(1)_F$ family charges when
the VEV's for the scalar fields and their conjugates become unequal.
The first or third family squark and slepton masses will be split further 
away
from the universal values than the second family, with the family 
receiving the 
smallest mass depending upon the sign of the splitting parameter present 
in 
(6.6).  Although the model discussed is far from the usual
minimal model, since only one pair of Higgs doublets survives at the 
electroweak scale, the scalar potential for the Higgs doublets at that 
scale 
is similar to that of the minimal supersymmetric standard model, MSSM.  
As 
such, the good result for $\sin^2 \theta_W$ is maintained.

As a result of the additional Higgs supermultiplets introduced in the 
model for the Higgs sector, several new terms appear in the Yukawa 
superpotential involving an extra conjugate pair of Higgs singlets and 
two 
${\bf 210}$ representations.  If their corresponding Yukawa couplings are 
not taken to vanish, they can alter the numerical results obtained 
earlier 
in Ref. \cite{an2}.  We shall defer for future study this point and the 
possible role the added neutrino singlets may play as sterile neutrinos
in neutrino oscillations.

Due to the complexity of the model, we have not checked that the 
potential 
can be minimized with VEV's that successfully break the $SO(10) \times 
U(1)_F$
symmetry as required, while all the conditions are satisfied that ensure
supersymmetry remains unbroken at the SUSY-GUT scale, just one pair of
Higgs doublets remains light and all Higgs triplets become massive.  As such,
we have only explored some of the issues that arise and have not succeeded
in building a completely self-consistent supersymmetric $SO(10) \times 
U(1)_F$ 
model.  But we have found some of the features of the model sufficiently 
interesting to report them here at this stage.
\vspace{0.2in}
\begin{center}
        ACKNOWLEDGMENTS
\end{center}

This research was carried out while the author was on sabbatical leave 
from Northern Illinois University.  He wishes to thank the Particle 
Physics 
groups at DESY, Fermilab, the University of Lund, the Werner Heisenberg
Institute of the Max Planck Institute, and the Technical University 
of Munich for their kind hospitality and financial support during this
period.  He is grateful to Joseph Lykken and Satya Nandi for their helpful
suggestions and advice.  Others he also wishes to thank for helpful 
comments and encouragement during the early stages of this research are 
Savas 
Dimopoulos, Emilian Dudas, Ralf Hempfling, Pham Q. Hung, Alex Kagan, Hans 
Peter Nilles, Stefan Pokorski and Peter Zerwas.  This work was supported 
in 
part by the U.S. Department of Energy.

\end{document}